# Multi-objective Test Case Selection Through Linkage Learning-based Crossover

Mitchell Olsthoorn[0000−0003−0551−6690] and Annibale Panichella[0000−0002−7395−3588]

Delft University of Technology, Delft, The Netherlands
M.J.G.Olsthoorn@tudelft.nl, A.Panichella@tudelft.nl

**Abstract.** Test Case Selection (`TCS`) aims to select a subset of the test suite to run for regression testing. The selection is typically based on past coverage and execution cost data. Researchers have successfully used multi-objective evolutionary algorithms (`MOEAs`), such as `NSGA-II` and its variants, to solve this problem. These `MOEAs` use traditional crossover operators to create new candidate solutions through genetic recombination. Recent studies in numerical optimization have shown that better recombinations can be made using machine learning, in particular linkage learning. Inspired by these recent advances in this field, we propose a new variant of `NSGA-II`, called `L2-NSGA`, that uses linkage learning to optimize test case selection. In particular, we use an unsupervised clustering algorithm to infer promising patterns among the solutions (subset of test suites). Then, these patterns are used in the next iterations of `L2-NSGA` to create solutions that preserve these inferred patterns. Our results show that our customizations make `NSGA-II` more effective for test case selection. The test suite sub-sets generated by `L2-NSGA` are less expensive and detect more faults than those generated by `MOEAs` used in the literature for regression testing.

## 1 Introduction

Software testing is one of the main phases in the software development life cycle. Developers write test cases for newly developed functionalities and maintain and update the existing test base. Regression testing aims to assess that changes to the production code do not affect the behavior of unchanged parts [27]. Ideally, regression testing can be tackled by running the entire test suite within a DevOps pipeline [24]. However, this strategy becomes unfeasible for very large systems in terms of resources (*e.g.,* build servers) and time. For this reason, researchers in the software engineering community proposed various techniques to reduce the cost of regression testing by removing redundant tests (*test suite minimization* [20]); or sorting the test cases with the goal of detecting regression faults earlier (*test case prioritization* [19]); or selecting fewer tests to run (*test case selection* [2]).

Multi-objective Evolutionary Algorithms (`MOEAs`) (and `NSGA-II` in particular) have been successfully used in the literature to produce Pareto efficient subsets of the test suites *w.r.t.* different testing criteria [27–29]. `MOEAs` that rely on Pareto ranking and problem decompositions have been shown to achieve good performance also compared to greedy algorithms and local solvers [29]. Further studies tailored the individual elements of `MOEAs` for test case selection, such as the initialization phase (*e.g.,* [14, 21, 30]) and selection operators [17].

One limitation for classic `MOEAs` (including `NSGA-II`) is that new solutions are generated using fully randomized recombination (crossover) operators [22, 26]. For example, the single-point and the multi-point crossovers randomly cut the chromosomes and exchange the genetic materials between two parent solutions, potentially breaking "promising" patterns. The latest advances in the evolutionary computation literature showed that a more effective search could be performed by identifying and preserving *linkage structures*, *i.e.,* groups of genes (problem variables) that should be replicated altogether into the offspring. *Linkage learning* [26] is a broad umbrella of methods to infer linkage structures and exploit this knowledge within more "competent" variation operators [16].

While *linkage learning* has been shown to be effective for single-objective numerical problems [16, 22, 26], we argue that it can also have huge potential for multi-objective test case selection. In this context, a solution is a binary vector where each bit $i$ indicates whether the $i$-th test case is selected or not for regression testing. `MOEAs` can generate partial solutions that contain promising patterns, *i.e.,* groups of test cases (bits) that together allow achieving high coverage with minimal execution cost. Hence, detecting and preserving these patterns (group of bits) using linkage learning can improve overall the search process.

This paper introduces `L2-NSGA`, a variant of `NSGA-II` that integrates key elements of *linkage learning* for the test case selection problem. In particular, `L2-NSGA` uses Agglomerative Hierarchical Clustering (AHC) to identify linkage structures in the non-dominated solutions produced by `NSGA-II` in every other generation. These structures (patterns) are groups of bits (subsets of test suites) that are found to be statistically frequent within the non-dominated solutions according to the AHC algorithm. Then, `L2-NSGA` uses a novel crossover operator that stochastically selects and replicates some of the inferred structures into new individuals. Given the multi-objective nature of regression testing, `L2-NSGA` optimizes all testing criteria simultaneously by using the *fast non-dominated sorting* algorithm and the *crowding distance* defined by Deb *et al.* for `NSGA-II` [3].

We conduct an empirical study on four software systems with multiple versions and regression faults. We analyze the quality and fault detection capability of the solutions produced by `L2-NSGA`. We compare its performance against `NSGA-II`, which is the most widely-employed MOEA in the regression testing literature (*e.g.,* [14, 21, 28, 30]). Our results suggest that the sub-test suites produced by `L2-NSGA` achieve higher coverage while incurring lower execution costs than the baseline (RQ1). Furthermore, the solutions created by `L2-NSGA` detect more regression faults than the solutions produced by `NSGA-II` (RQ2).

## 2  Background and Related Work

Three main approaches have been proposed in the literature to reduce the cost of regression testing [27]: *test suite minimization* [20], *test case prioritization* [19], and *test case selection* (TCS) [2]. Test case minimization aims to reduce the size of the test suite (number of test cases) by removing test cases that are redundant according to the chosen test criteria. Test case prioritization prioritizes (sorts) the test cases with the goal of running the fault-detecting tests earlier. Finally, TCS aims to select a subset of the original test suite taking into account test software changes and balancing cost (*e.g.,* execution time and resource usage) and test quality (*e.g.,* branch coverage). Given the conflicting nature of test quality and test resources, TCS is inherently a *multi-objective* problem [29] and, therefore, addressed using multi-objective evolutionary algorithms (MOEAs).

A common practice to decide which tests to select consists of using *test quality* metrics (or adequacy criteria) as surrogates for fault detection capability. These quality metrics reflect different aspects that software testers might be interested in maximizing, such as running test cases that exercise most of the production code as possible (code coverage [27, 29, 30]) or test cases that cover certain requirements first (requirement coverage [30]).

Yoo and Harman [29] introduced the first explicit formulation of TCS using a multi-objective paradigm. Given a program $P$, its new version $P'$, and a test suite $T$, the goal is to find Pareto efficient subset of test suites $T' \subseteq T$ that correspond to optimal trade-offs among the different testing criteria to optimize.

Let $\mathcal{Q} = \{Q_1, \ldots, Q_m\}$ *be the set of quality metrics to maximize; and let $C$ be the cost metric to minimize, the multi-objective TCS problem can be formulated using the following formula*:

$$min\ \Omega(T') = [C(T'), -Q_1(T'), \ldots, -Q_1(T')] \qquad (1)$$

*where $T' \subseteq T$; $Q_i(T)$ denotes the quality value of $T'$ based on the metric $Q_i$; and $C(T')$ is the cost of the sub-set $T'$.*

Solutions to the TCS are encoded as *binary chromosomes*, where the $i$-th binary element (or *gene*) is set to 1 if the test case $t_i \in T$ is selected; 0 otherwise. A solution $T_x \subseteq T$ is said to *dominate* another solution $T_y \subseteq T$ (denoted by $T_x \succ T_y$), if $T_x$ is better than or equal to $T_y$ for all test objectives, and there is at least one objective (*e.g.,* test cost) in which $T_x \subseteq T$ is strictly better than $T_y$. A solution $T^*$ is *Pareto optimal* if there exists no other solution $T_x \subseteq T$ such that $T_x \succ T^*$. The set of all Pareto optimal solutions (subsets of $T$) is called *Pareto optimal set*, and the corresponding objective vectors form the *Pareto front*. The goal of multi-objective TCS is to find Pareto optimal (or efficient) subsets of the test suite $T$ to run for regression testing.

Yoo and Harman [29] introduced multi-objective variants of the greedy algorithms for the set cover problem. They also assessed the performances of MOEAs, and NSGA-II in particular. Epitropakis *et al.* [4] empirically showed that MOEAs produce more effective solutions (*i.e.,* detect more regression faults) than greedy approaches. Yoo *et al.* [32] successfully applied MOEAs to reduce the cost of re-gression testing within Google's test environment. We observe that the most commonly used `MOEAs` for `TCS` is `NSGA-II` (*e.g.,* [12, 29–31]).

### 2.1   Linkage Learning

Evolutionary Algorithm (EAs) with simple *variation operators* have been shown to perform poorly for combinatorial problems with a high number of variable problems [8, 11]. This is because the effectiveness of EAs strongly depends on their ability to mix and preserve good partial solutions [10]. Prior studies proposed more effective variation operators (*i.e.,* crossover and mutation) that exploit *linkage information* to improve the scalability of EAs [15,18]. Linkage information can be inferred using different techniques, such as Bayesian Network [16], Dependency Structure Matrix [33], and hierarchical clustering [22, 23].

Gene-pool Optimal Mixing Evolutionary Algorithm (`GOMEA`) is one of the latest linkage-based evolutionary algorithm. `GOMEA` uses hierarchical clustering to learn linkage tree structures and uses the linkage information to create new solutions. In particular, `GOMEA` uses *gene optimal mixing* to improve existing solutions iteratively using local search and evaluating partial solutions. In each generation, `GOMEA` infers the linkage tree using the UPGMA algorithm [23]. UPGMA is a bottom-up approach that clusters genes (problem variables) based on their similarities. The similarity is computed using the normalized *mutual information*. The result of UPGMA is a linkage structure, called *Family Of Subsets* (FOS). A FOS is a set $\{F^0, F^1, \ldots, F^{|F|-1}\}$ where each $F^i$ is a subset of the gene indexes. For example, the set $F^i = \{1, 2, 3\}$ indicates that the genes at index 1, 2, and 3 are linked together and should be considered a unique "block". The family of subsets (FOS) is a tree that contains $N$ leaf nodes and $N-1$ internal node, where $N$ is the number of variable problems. The leaf nodes correspond individual variable problems (univariate subsets), while the internal nodes merge the child nodes into larger subsets.

Given the FOS, `GOMEA` creates new solution by iteratively applying each subset (linkage structure) into a parent individual and accepting only changes that strictly improve the fitness function. In other words, `GOMEA` applies an exhaustive local search on the linkage structures. Although `GOMEA` is very effective at achieving better convergence for problems with large numbers of decision variables, it is not designed for multi-objective problems such as `TCS`. In our context, we have multiple conflicting objectives (testing criteria) that must be optimized simultaneously. Another limitation of `GOMEA` is that it uses a computationally expensive local search heuristic to try all possible linkage structures through many fitness evaluations.

## 3   Approach

In this paper, we introduce a variant of `NSGA-II` that incorporates key elements of `GOMEA`, hereafter referred to as `L2-NSGA`. Algorithm 1 outlines the pseudo code of `L2-NSGA`. The algorithm starts with an initial pool of random solutions,

**Algorithm 1:** `L2-NSGA`

```
 1 begin
 2     P ⟵ INITIAL-POPULATION()
 3     while not (end condition) do
 4         FOS ⟵ INFER-MODEL(P, 2)
 5         P' ⟵ ∅
 6         forall i in 1..|P| do
 7             Parent ⟵ TOURNAMENT-SELECTION(P)
 8             Donor ⟵ TOURNAMENT-SELECTION(P)
 9             Child ⟵ L2-CROSSOVER(Parent, Donor, FOS)
10             Child ⟵ MUTATE(Child)
11             P' ⟵ P' ⋃ {Child}
12         R ⟵ P' ⋃ P
13         𝔽 ⟵ FAST-NONDOMINATED-SORT(R)
14         P ⟵ ∅
15         d ⟵ 1
16         while |P| + |𝔽_d| ≤ M do
17             CROWDING-DISTANCE-ASSIGNMENT(𝔽_d)
18             P ⟵ P ⋃ 𝔽_d
19             d ⟵ d + 1
20         SORT-BY-CROWDING-DISTANCE(𝔽_d)
21         P ⟵ P ⋃ 𝔽_d[1 : (M − |P|)]
22     return 𝔽_1
```

*i.e.,* random subsets of test suites (line 2). The population then evolves through subsequent generations to find nearby non-dominated solutions (loop in 3-24). In line 4, the algorithm infers the linkage structures from the best individuals in the population $P$ using UPGMA. The structures (FOS) are only inferred every other iteration to reduce the overhead of the inference process. To produce the population for the next generation, `L2-NSGA` first creates new individuals, using the L2-CROSSOVER, which uses the model inferred in line 4. We describe the crossover operator in detail in Section 3.1. In particular, the *binary tournament* selection is applied to select two individuals from the population: one parent (line 7) and one donor (line 8). In lines 9-10, the child solution is created by applying our new crossover operator (Section 3.1) and the *bit-flip mutation* operator. Once the offspring population $P'$ is obtained, `L2-NSGA` uses the *fast non-dominated sorting algorithm* and *crowding distance* from `NSGA-II` to form the population for the next generation (*elitism*).

In line 13, the parent ($P$) and the offspring ($P'$) populations are combined into one single pool $R$. The solutions in $R$ are ranked in subsequent *non-dominated fronts* using the *fast non-dominated sorting* algorithm by Deb *et al.* [3]. The solutions in the first front $\mathbb{F}_1 \subseteq R$ are not dominated by any other solution in $P$; the solutions in the second front $\mathbb{F}_2$ are dominated by the solutions in $\mathbb{F}_1 \subset R$ but do not dominate one another; and so on.

The loop between lines 17 and 21 adds as many individuals to the next generation as possible, based on their non-dominance ranks, until reaching the population size. `L2-NSGA` first selects the non-dominated solutions from $\mathbb{F}_1$; if the number of selected solutions is lower than the population size, the loop selects the non-dominated solutions from $\mathbb{F}_2$, etc. The loop ends when adding the solutions of the current front $F_d$ exceeds the maximum population size (the condition in line 17). In the latter case, the algorithm selects the remaining solutions from the front $\mathbb{F}_d$ according to the descending order of *crowding distance* in lines 22-23.

Notice that the *binary tournament selection* selects parent and donor solutions using the concept of Pareto optimality, which leads to selecting individuals with better (lower) dominance ranks. Further, the *crowding distance* increases the selection probability for the more diverse individuals within the same non-dominance front. The main loop in lines 3-24 terminates when the maximum number of generations is reached.

### 3.1  Linkage-based Crossover

Given one *parent* solution and one *donor* solution, our goal is to copy genes from the donor to the parent using the linkage structures (FOS). Let FOS = $\{F^0, \ldots, F^n\}$ be the *family of subsets* produced the UPGMA on the population $P$. The new solution `Child` is obtained by cloning the parent solution and replicating $K$ randomly selected subsets $F^j \in$ FOS from the donor solution. More formally, let $\text{FOS}_K \subset \text{FOS}$ be the set of $K$ subsets randomly chosen from FOS; the new solution `Child` is formed as follows:

$$\texttt{Child}[i] = \begin{cases} \texttt{Parent}[i] & i \notin \text{one of the sets in FOS}_K \\ \texttt{Donor}[i] & i \in \text{one of the sets in FOS}_K \end{cases} \quad (2)$$

where `Child`$[i]$ is the i-th gene of the child solution; `Parent`$[i]$ is the i-th gene the parent solution; and `Donor`$[i]$ is the i-th gene of the donor solutions.

The donor genes are replicated into the child altogether without applying the computationally expensive local search of `GOMEA`. Another important difference compared to `GOMEA` is that `L2-NSGA` always accepts the child solutions; parents and offsprings are selected for the next generation based on their dominance ranks and crowding distance values (lines 17-23 in Algorithm 1). Instead, `GOMEA` iteratively accepts partial changes only if they do not worsen the current single fitness value. Recall that `GOMEA` is a single-objective search algorithm.

In our preliminary experiments, we assessed different $K$ values for the number of FOS to copy into the parent. We obtained good results for the systems used in our empirical study when setting $K$ equal to 50 % of the linkage structures.

### 3.2  Similarity Function for Linkage Learning

UPGMA is a bottom-up iterative algorithm for hierarchical clustering, and it is used in both `GOMEA` and `L2-NSGA` to infer linkage tree structures. UPGMA requires to choose a distance function $d$ to compute the linkage tree, *i.e.,* to decide which subsets to merge in each iteration. Various distance functions can be used, such as the *mutual information* [9], [23], *hamming distance*, and the *correlation coefficient*. `L2-NSGA` uses the *hamming distance* as it the classical distance function for binary vectors (*i.e.,* solution in our case) and has a much lower computational complexity. The hamming distance between two problem variables $X$ and $Y$ corresponds to the number of substitutions to apply to $X$ to obtain $Y$. The computational complexity of the hamming distance is $O(N \times M)$, with $N$ being the population size, and $M$ being the length of the chromosomes.

Table 1: Programs used in the study.

| Program | Versions | LOC | # Tests | Fault Type | Language |
|---|---|---|---|---|---|
| bash | {v1, v2, v3} | 44,991 – 46,294 | 1,061 | Seeded | C |
| flex | {v1, v2, v3} | 9,484 – 10,243 | 567 | Seeded | C |
| grep | {v1, v2, v3} | 9,400 – 10,066 | 806 | Seeded | C |
| sed  | {v1, v2, v3} | 5,488 – 7,082 | 360 | Seeded | C |

## 4 Empirical Study

We formulated the following research questions:

**RQ1** *To what extent does `L2-NSGA` produce better Pareto efficient solutions compared to `NSGA-II`?*

**RQ2** *What is the cost-effectiveness of the solution produced by `L2-NSGA` vs. `NSGA-II`?*

In particular, we assess the performance of `L2-NSGA` by comparing it with `NSGA-II` (the baseline). `NSGA-II`, which is the most frequently used `MOEAs` in regression testing [4, 14, 21, 29, 30]. `NSGA-II` is also a logical baseline since our approach extends it with linkage learning. RQ1 aims to assess to what extent `L2-NSGA` produces better solutions (subsets of test suites) than `NSGA-II` with regards to given test adequacy and cost criteria. RQ2 aims to understand how many faults can be detected by the solutions produced by the two `MOEAs`. This research question reflects practitioners' needs, interested in reducing the cost of regression testing without reducing the number of detected regression faults.

**Benchmark**. Our empirical study includes multiple versions of four real-world programs written in `C`: `bash`, `flex`, `grep`, and `sed`. Table 1 summarizes the main characteristics (*e.g.,* test suite size and version) of the programs in our benchmark. These programs are selected from the *Software-artifact Infrastructure Repository* (SIR) [6]. The four programs correspond to the well-known UNIX utilities obtained from the GNU website and are the largest programs written in `C` available in SIR. SIR provides multiple subsequent versions of the programs and their test suites. The test suite size varies from 360 to 1061 test cases, which have been created by applying both white-box (statement coverage) and black-box (category partition) test adequacy criteria [6]. SIR also includes faulty versions of these programs with seeded (artificial) faults. Similar to what has been done in the literature (*e.g.,* [29]), we considered non-trivial seeded faults that only few tests can detect. It is worth noting that these UNIX utilities have been widely used in the related literature on regression testing (*e.g.,* [4,7,29,30]).

**Test Objectives**. In our study, we considered three test criteria, which corresponds to the three objectives to optimize. In particular, we considered *statement coverage* (to maximize), *branch coverage* (to maximize), and *execution cost* (to minimize). For each available test case $T$, we stored the code branches and statements that $T$ covers using `gcov`, which is a coverage tool from the `C` compiler in GNU. To have a reliable measure of test execution time, we counted the number of instructions executed by each test case $T$ by using the `gcov` tool. This methodology of measuring the test execution cost has been widely used in the related literature to avoid biased measurements due to both the hardware and software environments used to run the test suites [29, 30].

**Experimental Protocol**. We run the two `MOEAs` 20 times (each) to address their randomized nature. In each run, we stored the non-dominated front and the corresponding optimal solution set produce at the end of the search budget, *i.e.,* when reaching the maximum number of evaluated solutions.

To answer RQ1, we need to compare the non-dominated fronts produced by the `MOEAs` with regard to the optimal (true) Pareto front. Since the `TCS` problem is NP-complete [5], it is not possible to compute the true Pareto front in polynomial time for the programs in our benchmark. Hence, we build the so-called *reference front* [29, 30] that combines the best parts of the non-dominated fronts produced by both `NSGA-II` and `L2-NSGA` across all 20 runs. More precisely, let $\{F_1, \ldots, F_k\}$ be all non-dominated fronts produced by `L2-NSGA` and `NSGA-II`; the reference front $\mathcal{R}$ is built as follows:

$$\mathcal{R} \subseteq \bigcup_{i=1}^{k} F_i, \forall p_1 \in \mathcal{R}, p_2 \in \mathcal{R} : p_2 \prec p_1 \tag{3}$$

Given the *reference front* $\mathcal{R}$, we then used the *inverted generational distance* (IGD) and the *hypervolume* (HV) as the *quality indicators* [34]. IGD measures both proximity of the non-dominated fronts produced by `MOEAs` to $\mathcal{R}$ as well as the solution diversity [34]. Therefore, smaller IGD values are preferable. In contrast, HV measures the area/volume that is dominated by a non-dominated front. Hence, larger HV values are preferable.

To answer RQ2, we analyze the fault detection capabilities of the subset of test suites produced by `L2-NSGA` and `NSGA-II`. To this aim, we analyze the number of regression faults that can be detected by each solution (sub-suite) in a given non-dominated front $F_i$. Hence, a non-dominated front $F_i$ produced by a MOEA corresponds to a set of points with different cost values and different number of detected faults (cost-effective front). Non-dominated fronts that detects more faults with lower execution cost are preferable.

To quantify the *cost-effectiveness* into a single scalar value, we used the normalized Area Under the Curve (AUC) delimited by the cost-effective front ($I_{CE}$ metric). This metric has been used to measure the average-fault detection capability of a non-dominated front for multi-objective `TCS` approaches [14]. $I_{CE}$ ranges within the interval $[0; 1]$. $I_{CE} = 1$ corresponds to the ideal (utopia) case where the front $F_i$ detects all faults independently on the number of selected tests. $I_{CE} = 0$ indicates the worst-case scenario where no faults are detected. Hence, larger $I_{CE}$ values are better.

To assess the significance of the differences among `L2-NSGA` and `NSGA-II`, we use the Wilcoxon rank-sum test with the threshold $\alpha$=0.05. A significant *p*-value indicates that `L2-NSGA` achieves better performance metric compared to `NSGA-II` across 20 runs. We further complement our statistical analysis with the Vargha-Delaney statistic ($\hat{A}_{12}$) to measure the effect size of the results. $\hat{A}_{12} > 0.5$ indicates a positive effect size for `L2-NSGA`.

**Parameter Setting**. For our experiments, we used the same parameter values used in the literature [29, 30]. In particular, both `NSGA-II` and `L2-NSGA` use a *population* size of 100 solutions; and the *bit-flip mutation* operator with probability $p_m = 1/n$, with $n$ being the test suite size (*i.e.,* chromosome length). For the recombination operators, `NSGA-II` uses the *scattered crossover* [14] with the probability $p_c = 0.8$. Instead, `L2-NSGA` uses the L2-CROSSOVER presented in Section 3 with the probability $p_c = 0.8$ as well. For both `NSGA-II` and `L2-NSGA`, we use the *binary tournament* selection. Finally, we used the stopping condition of 20 000 fitness evaluations, or equivalently 200 generations.

**Implementation**. We implemented `L2-NSGA` in Python using the `pymoo` library [1]. We use the implementation of linkage leaning and hierarchical clustering available in `SciPy`. The source code of `L2-NSGA` is publicly available as an artifact [13]. In our implementation, we pre-processed coverage data using the lossless *compaction algorithm* by Epitropakis *et al.* [4]. This algorithm has been proved to reduce the fitness evaluation cost drastically.

## 5 Results

In this section, we discuss the results achieved for each research question separately. Section 6 elaborates on the potential threats to the validity of our study.

**Results for RQ1**. Table 2 reports the median and IQR (interquartile range) of the IGD and HV values achieved by `L2-NSGA` and `NSGA-II`. The IGD metric has been computed using the *reference front* built as described in the study design. We observe that `L2-NSGA` achieves smaller (better) IGD values than the baseline. This means that the subset of test suites produced by `L2-NSGA` are much closer to the reference front and more well-distributed compared to `NSGA-II`. We further observe that the HV scores achieved by `L2-NSGA` are always larger than those obtained with `NSGA-II`. This means that the non-dominated fronts produced by our approach are closer to the reference fronts (IGD) and dominate a larger portion of the objective space (HV).

To provide a graphical interpretation of the results, Fig. 1 depicts the non-dominated fronts produced by the two `MOEAs` for `sed` version `v1` when using two different numbers of generations. For the sake of this analysis, we selected the front achieving the median IGD value across the 20 independent runs. Fig. 1a shows the front produced by the two `MOEAs` after 100 generations. `L2-NSGA` produced a better distributed front compared to `NSGA-II`. Besides, the solutions by `L2-NSGA` dominate all the solutions produced by the baseline. This means that developers can choose sub-suites from `L2-NSGA` that yield the same or larger coverage but incur a much smaller test execution cost. Fig. 1b shows the front produced by the two `MOEAs` for the same system, but after 200 generations. Also in this case, `L2-NSGA` produced a better distributed front than `NSGA-II`.

Table 2: Median IGD and HV values (with IQR) achieved by the `L2-NSGA` and `NSGA-II`. Best performance is highlighted in grey color.

| System Version | | IGD | | HV | |
|---|---|---|---|---|---|
| | | NSGA-II | L2-NSGA | NSGA-II | L2-NSGA |
| bash | v1 | 0.1987 (0.0192) | 0.1046 (0.0224) | 0.4165 (0.0276) | 0.6418 (0.0483) |
| | v2 | 0.2059 (0.0238) | 0.1136 (0.0201) | 0.6223 (0.0215) | 0.7710 (0.04242) |
| | v3 | 0.2839 (0.0333) | 0.1221 (0.0300) | 0.3638 (0.0252) | 0.6110 (0.0435) |
| flex | v1 | 0.0300 (0.0068) | 0.0265 (0.0058) | 0.9924 (0.0014) | 0.9937 (0.0016) |
| | v2 | 0.0324 (0.0144) | 0.0230 (0.0086) | 0.9810 (0.0038) | 0.9853 (0.0060) |
| | v3 | 0.0519 (0.0333) | 0.0350 (0.0159) | 0.9808 (0.0053) | 0.9857 (0.0028) |
| grep | v1 | 0.1872 (0.0881) | 0.0995 (0.0467) | 0.4623 (0.0908) | 0.6327 (0.0983) |
| | v2 | 0.1702 (0.0573) | 0.1301 (0.0865) | 0.5246 (0.0782) | 0.5991 (0.1575) |
| | v3 | 0.1920 (0.0835) | 0.1428 (0.0389) | 0.4310 (0.0851) | 0.5540 (0.0968) |
| sed | v1 | 0.1123 (0.0834) | 0.0544 (0.0570) | 0.8863 (0.06489) | 0.9580 (0.0552) |
| | v2 | 0.0546 (0.0141) | 0.0158 (0.0258) | 0.9508 (0.0393) | 0.9900 (0.0209) |
| | v3 | 0.0752 (0.0617) | 0.0253 (0.0143) | 0.8919 (0.0805) | 0.9761 (0.0433) |

Table 3 reports the statistical test results, namely the *p*-value of the Wilcoxon rank-sum test and the $\hat{A}_{12}$ statistic. According to these tests, the differences between the two `MOEAs` are statistically significant (*p*-value $< 0.01$) for all systems and all versions. The only exception to this rule is `flex v1` for which the *p*-value is only marginally significant when considering the IGD metric. The effect size is always *large* for HV. These results suggest that `L2-NSGA` achieves better results independently of the size of the project and the test suites.

**Results for RQ2**. Table 4 reports the median and IQR of the $I_{CE}$ values achieved by `L2-NSGA` and `NSGA-II`. Recall that $I_{CE}$ measures the average number regression faults detected by `MOEAs` at different execution cost intervals (see Section 4 for more details). We observe that `L2-NSGA` achieves better (larger) $I_{CE}$ values than `NSGA-II` in 10 out of 12 comparisons. Our approach achieves better results for all versions of `bash`, `grep`, and `sed`. The difference ranges from +19p.p. (percentage points) for `bash v3` to +0.10p.p. for `flex v3`. Instead, for `flex v1` and `v2`, the two approaches provide almost identical $I_{CE}$ scores.

According to the Wilcoxon rank-sum tests and the Vargha-Delaney statistic (results also reported in Table 4), `L2-NSGA` statistically outperforms the baseline *w.r.t.* cost-effectiveness in all versions of `bash`, `grep`, and `sed` with a *large* effect size. For `flex`, the statistical significance only holds for version `v3` with a *large* effect size. Lastly, there is no statistical difference between the two `MOEAs` for versions `v1` and `v2` (negligible effect size).

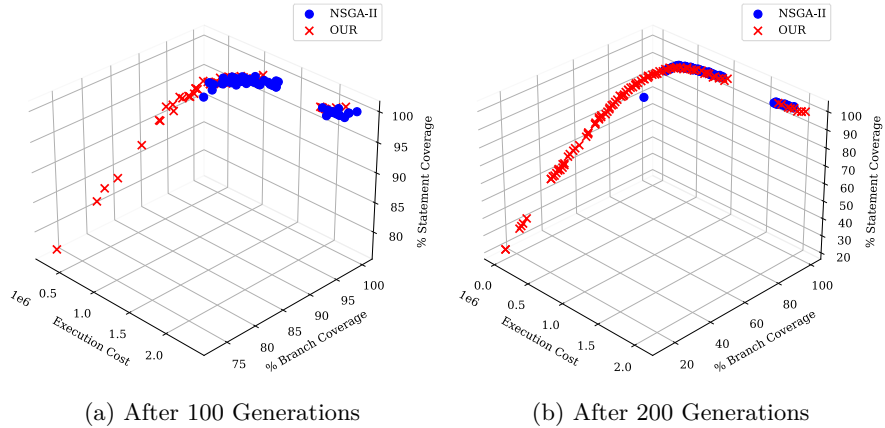

(a) After 100 Generations     (b) After 200 Generations

Fig. 1: Fronts produced by `L2-NSGA` and `NSGA-II` for `sed` version `v1`

Table 3: Results of the statistical tests

| **System** | **Version** | **IGD** | | **HV** | |
|---|---|---|---|---|---|
| | | $p$-value | $\hat{A}_{12}$ | $p$-value | $\hat{A}_{12}$ |
| bash | v1 | <0.01 | 1.00 (large) | <0.01 | 1.00 (large) |
| | v2 | <0.01 | 1.00 (large) | <0.01 | 1.00 (large) |
| | v3 | <0.01 | 1.00 (large) | <0.01 | 1.00 (large) |
| flex | v1 | 0.06 | 0.64 (small) | <0.01 | 0.77 (large) |
| | v2 | 0.04 | 0.66 (small) | <0.01 | 0.80 (large) |
| | v3 | <0.01 | 0.72 (med.) | <0.01 | 0.85 (large) |
| grep | v1 | <0.01 | 0.88 (large) | <0.01 | 0.94 (large) |
| | v2 | <0.01 | 0.76 (large) | <0.01 | 0.805 (large) |
| | v3 | <0.01 | 0.81 (large) | <0.01 | 0.95 (large) |
| sed | v1 | <0.01 | 0.76 (large) | <0.01 | 0.83 (large) |
| | v2 | <0.01 | 0.91 (large) | <0.01 | 0.92 (large) |
| | v3 | <0.01 | 0.87 (large) | <0.01 | 0.91 (large) |

**Running Time Analysis**. Compared to `NSGA-II`, our approach applies linkage learning every two generations. The inference is based on UPGMA, which is a known fast algorithm for clustering. Nevertheless, this algorithm adds some extra overhead to the search process. Hence, it is important to quantify such overhead for practical purposes. To this aim, Fig. 2 reports the execution time spent by each algorithm to perform 200 generations on each program and independent run. The execution time was measured using a machine with Intel Core i7 processor running at 2.40GHz with 16GB RAM.

From Fig. 2, we observe that `L2-NSGA` is, on average, 30 % slower than `NSGA-II` for all programs (and versions) in our benchmark. For the two smallest programs (*i.e.,* `flex` and `sed`) the differences are between +1 s and +2 s. Hence, the differences are very negligible in practice. A larger overhead is noticeable for `bash` and `grep`, which have the largest test suites. For the former system, the differences in running time between L2-NSGA and NSGA-II are between $+10$ s and $+15$ s. For the latter, our approach requires between $+5$ s and $+6$ s compared to the baseline. However, we can conclude that a difference of a few seconds are very negligible in a practical setting.

Table 4: Cost-effective results. Best performance is highlighted in grey color.

| System | Version | NSGA-II $I_{CE}$ | IQR | L2-NSGA $I_{CE}$ | IQR | Stat. Analysis $p$-value | $\hat{A}_{12}$ |
|---|---|---|---|---|---|---|---|
| bash | v1 | 0.6857 | 0.0171 | 0.8566 | 0.0395 | <0.01 | 1.00 (large) |
|  | v2 | 0.5711 | 0.0586 | 0.7031 | 0.0981 | <0.01 | 0.98 (large) |
|  | v3 | 0.6760 | 0.0770 | 0.8559 | 0.0677 | <0.01 | 0.96 (large) |
| flex | v1 | 0.6718 | 0.0476 | 0.6721 | 0.0478 | 0.31 | 0.55 (negl.) |
|  | v2 | 0.5243 | 0.0008 | 0.5244 | 0.0248 | 0.51 | 0.50 (negl.) |
|  | v3 | 0.6809 | 0.0018 | 0.6827 | 0.0009 | <0.01 | 0.89 (large) |
| grep | v1 | 0.3725 | 0.0323 | 0.43031 | 0.0586 | <0.01 | 0.83 (large) |
|  | v2 | 0.3474 | 0.0325 | 0.4260 | 0.0452 | <0.01 | 0.92 (large) |
|  | v3 | 0.1370 | 0.0088 | 0.2052 | 0.0214 | <0.01 | 1.00 (large) |
| sed | v1 | 0.7552 | 0.0214 | 0.7760 | 0.0178 | <0.01 | 0.85 (large) |
|  | v2 | 0.9275 | 0.0350 | 0.9414 | 0.0340 | <0.01 | 0.74 (large) |
|  | v3 | 0.9476 | 0.0416 | 0.9894 | 0.0204 | <0.01 | 0.91 (large) |

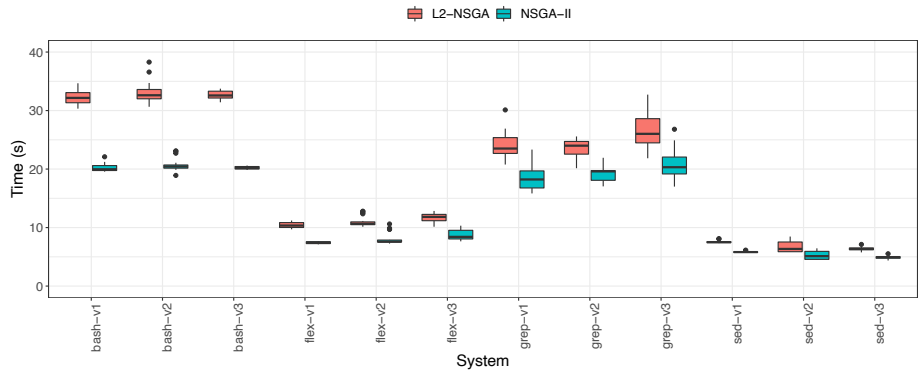

Fig. 2: Execution Time (in seconds) of NSGA-II and L2-NSGA

Lastly, we note that running NSGA-II for longer will not improve its results (IGD, HV, and $I_{CE}$). For example, running NSGA-II on bash v1 for 600 generations (for 80 s on average) leads to HV= 0.35 (compared to 0.42 of L2-NSGA in Table 2 with 200 generations) and an $I_{CE} = 0.69$ (compared to 0.86 of L2-NSGA in Table 4 with 200 generations). These are the median values achieved over 20 independent runs.

## 6    Threats to Validity

The threats to *construct validity* are related to the metrics we used to assess the MOEAs. In our study, we used the *inverted generation distance* (IGD) and the *hyper-volume* (HV) to answer our RQ1. These metrics are well-established quality indicators for multi-objective algorithms [34]. This is also in line with the guidelines by Wang *et al.* [25] that recommended using IGD for problems (like `TCS`) with no known reference points beforehand. We combined the non-dominated fronts produced by all `MOEAs` and across all runs to build the reference front. This is a standard practice in regression testing [14,21,29,30,32]. Another potential threat is related to the test quality and cost metrics we optimize with the `MOEAs`. To collect the coverage and test execution cost data, we relied on the `gcov` profiling tool as done in the literature [4,30].

Threats to *internal validity* are related to the random nature of `MOEAs` and the L2-CROSSOVER operator. To address this threat, we run each MOEA 20 times on each program version. Then, we analyze the median results and rely on non-parametric statistical tests (*i.e.,* Wilcoxon test and the Vargha-Delaney statistics) to draw our conclusions.

Threats to *external validity* regard the generalizability of our results. We selected four medium to large size software systems written in C. These systems are well-known UNIX utilities from the SIR dataset [6] and have been widely used in prior studies in regression testing [14,21,28,30]. Furthermore, we considered three different versions of each program in the benchmark. Replicating our study with more programs, further releases, and more `MOEAs` is part of our future plan.

## 7  Conclusions and Future Work

In this paper, we have introduced a novel approach, called `L2-NSGA`, for multi-objective test case selection (`TCS`). `L2-NSGA` extends `NSGA-II` by incorporating linkage learning methods. Inspired by the latest advances in evolutionary computation, our approach replaces the fully-randomized crossover operator of `NSGA-II` with a new operator (L2-CROSSOVER) that identifies, preserves, and replicates patterns of genes (bits) that characterize the fittest solutions in a given population. These patterns (also called linkage structures) are inferred through agglomerative hierarchical clustering and UPGMA in particular.

We evaluated `L2-NSGA` on four real-world programs with large test suites and multiple versions. Our results showed that `L2-NSGA` produces better non-dominated fronts than its predecessor `NSGA-II` (the baseline), widely used in the literature. Furthermore, the test suites created with `L2-NSGA` can detect more regression faults than the solutions produced by the baseline.

As future work, we plan to consider alternative clustering algorithms to learn the linkage structures and different stochastic approaches for selecting the genes to replicate into new solutions during recombination. We aim to assess the usefulness of `L2-NSGA` for other regression testing techniques, such as test case prioritization. Finally, we would like to combine linkage learning with other `MOEAs` widely used in search-based software testing.